\documentclass[showpacs,preprintnumbers,amsmath,amssymb]{revtex4}
\usepackage{graphicx}
\usepackage{dcolumn}
\makeatletter
\input{epsf}
\begin{document}
\title{ Statistics of Quasiparticles in Fractional Quantum Hall States}
\author{B. Basu}
 \email{banasri@isical.ac.in}

\author{P. Bandyopadhyay}
 \email{pratul@isical.ac.in}
\affiliation{Physics and Applied Mathematics Unit\\
 Indian Statistical Institute\\
 Kolkata-700108 }
\begin{abstract}
\begin{center}
Abstract
\end{center}
We have considered here the statistics parameter for
quasiparticles in FQH states from an analysis of these states in
the framework of chiral anomaly and Berry phase. It is shown that
we have  a generalized relation such that the statistical phase of
a quasiparticle for FQH states at  $\nu=p/q$ with $p(q)$ even or
odd, is given by $e^{i\pi \theta}$  with $\theta=p/q $ and the
charge is $-e \frac{p}{q}$. The statistics parameter for a
quasihole is identical with that of a quasiparticle and the charge
is of opposite sign.
\end{abstract}
\maketitle
\section{introduction}

In a seminal paper \cite{aro} Arovas, Schrieffer and Wilczek
derived the statistics of the quasiholes at $\nu=1/m$, $m$ odd in
a Berry phase calculation. They examined the Berry phase
corresponding to one quasihole encircling the origin and
interpreted this as the Aharanov-Bohm phase. The charge was found
to be $e/m$. Considering a pair of quasiholes encircling one
another they found the statistics parameter to have the value
$1/m$. The Laughlin quasielectrons were examined along the same
line \cite{geo} and within certain approximations, the charge and
statistics parameter were found to have the values $-e/m$ and
$1/m$ respectively.  Kjonsberg and Myrheim \cite{myr} have argued
that for FQH states the quasiparticles(qp) at $\nu=1/m$, $m$ odd,
do $not$ possess well defined statistics. Their numerical studies
show that at $\nu=1/3$ the charge and statistics parameter for
quasiholes(qh) are $e/3$ and $1/3$ respectively. However, for the
quasiparticles though a  slow convergence towards the expected
value of the charge $-e/3$ is observed with a finite size
correction for $N$ electrons, the statistics parameter has no well
defined value even for 200 electrons but might possibly converge
to $1/3$.

Johnson and Canright \cite{john} examined the exclusion statistics
parameter considering state counting based on numerical
simulations for interacting electrons on a sphere. It is found
that the one dimensional exclusion statistics parameter for
quasiholes is $1/m$ while for quasielectrons the value is $(2-
\frac{1}{m})$. The exclusion statistics parameter is the same
parameter as is obtained from  Berry phase analysis with an
opposite sign for the quasielectrons since their charge is
negative. This implies that the exclusion statistics parameter is
$1/m$ for quasiholes and $(-2+ \frac{1}{m})$ for quasielectrons.
Though the values $1/m$ and $-2 + \frac{1}{m}$ represent the same
particle statistics, their states are different, $1/m$ state
singular and $-2 + \frac{1}{m}$ state nonsingular.

In the composite fermion (CF) scenario Kjonsberg and Leinaas
\cite{lei} calculated the statistics of the $unprojected$ CF
quasiparticle at $\nu=1/m$, the wave function of which is
different from Laughlin function, and observed a definite value
though the sign is inconsistent with general considerations. Jeon,
Graham and Jain \cite{jeon} have confirmed that the statistics is
robust to projection into the lowest Landau level and argued that
the sign enigma has its origin in very small perturbations in the
trajectory due to the insertion of an additional CF quasiparticle.
These authors have numerically studied the statistics of the
composite fermion quasiparticles at $\nu=1/3$ and $\nu=2/5$ by
evaluating the Berry phase for a closed loop encircling another CF
quasiparticle and the statistics parameters were found to be
$-\frac{2}{3}$ and $-\frac{2}{5}$ respectively. The negative sign
has been attributed to the fact that there is an inter-CF
interaction which is weak and often attractive and involves a
significant overlap of CF quasiparticles causing small
perturbation in the trajectory.

Here we shall address the problem of statistics in a heuristic
manner on the basis of analysis of FQH states in the framework of
chiral anomaly and Berry phase. From a calculation of the Berry
phase of a quasiparticle (quasihole) encircling another
quasiparticle (quasihole) we shall show that the charge and
statistics parameter for quasiparticles at $\nu=\frac{n}{2mn \pm
1}$ with $m,n$ posotive integers are given by - $\nu e$ and
$\frac{n}{2mn \pm 1}$ respectively. For quasiholes the statistics
parameter is identical with that of the quasiparticle and the
charge has the opposite sign.

In sec. 2 we shall recapitulate certain features of the FQH states
analysed from the Berry phase approach. In sec. 3 we shall
calculate the statistics parameter for quasiparticles (quasiholes)
for FQH states at different filling factors.

\section{Fractional Quantum Hall States: A Berry Phase Approach}

In some earlier papers \cite{pb1,pb2} we have analyzed the
sequence of quantum Hall states from the viewpoint of chiral
anomaly and Berry phase. To this end, we have taken quantum Hall
states on the two dimensional surface of a $3D$ sphere with a
magnetic monopole of strength $\mu$ at the centre. In this
spherical geometry, we can analyze quantum Hall states in terms of
spinor wave functions and take advantage of the analysis in terms
of chiral anomaly which is associated with the Berry phase. In
this geometry the angular momentum relation is given by

\begin{equation}
{\bf J} ~ = ~ {\bf r} \times {\bf p} - \mu {\bf \hat{r}} ,
~~~~~~~~ \mu ~=~ 0 , ~\pm ~1 / 2 , ~\pm ~1 , ~\pm ~ 3 /
2,~.........
\end{equation}

From the description of spherical harmonics $Y^{m, \mu}_{\ell}$
with $\ell=1/2,~|m|=|\mu|=1/2$, we can construct a two-component
spinor $\theta = \left(
\begin{array}{c}
                          \displaystyle{u}\\
                          \displaystyle{v} \end{array} \right)$~~where
\begin{equation}
\begin{array}{lcl}
u ~=~ Y^{1/2 , 1/2}_{1/2} &=& \displaystyle{sin ~\frac{\theta}{2}
\exp
\left[ i (\phi - \chi) / 2 \right]}\\ \\
v ~=~ Y^{- 1/2 , 1/2}_{1/2} &=& \displaystyle{cos
~\frac{\theta}{2} \exp \left[- i (\phi + \chi) / 2 \right]}
\end{array}
\end{equation}
Here $\mu$ corresponds to the eigenvalue of the operator $i
\frac{\partial}{\partial \chi}$.

The $N$-particle wave function for the quantum Hall fluid state at
$\nu=\frac{1}{m}$ can be written as
\begin{equation}
{\psi^{(m)}}_{N} ~=~ \prod_{i < j} {(u_i v_j - u_j v_i)}^{m}
\end{equation}
$m$ being an odd integer. Here $u_i(v_j)$ corresponds to the
$i$-th ($j$-th) position of the particle in the system.

It is noted that ${\psi^{(m)}}_{N}$ is totally antisymmetric for
odd $m$ and symmetric for even $m$. We can identify  \cite{hal}
$m$ as $m=J_i+J_j$ for the N-particle system where $J_i$ is the
angular momentum of the $i$-th particle. It is evident from eqn.
(1) that with ${\bf r} \times {\bf p}=0$ and $\mu=\frac{1}{2}$ we
have
 $m=1$ which corresponds to  the complete
filling of the lowest Landau level.
 From the Dirac quantization condition $e
\mu =\frac{1}{2}$, we note that this state corresponds to $e=1$
describing the IQH state with $\nu=1$.

The next higher angular momentum state can be achieved either by
taking ${\bf r} \times {\bf p}=1$ and $\mid\mu \mid=\frac{1}{2}$
(which implies the higher Landau level) or by taking ${\bf r}
\times {\bf p}=0$ and $\mid{\mu_{eff}} \mid=\frac{3}{2}$ implying
the ground state for the Landau level. However, with
$\mid{\mu_{eff}} \mid=\frac{3}{2}$, we find the filling fraction
$\nu=\frac{1}{3}$ which follows from the condition $e \mu
=\frac{1}{2}$ for $\mu=\frac{3}{2}$. Generalizing this, we can
have $\nu=\frac{1}{5}$ with $\mid{\mu_{eff}} \mid=\frac{5}{2}$. In
this way we can explain all the FQH states with
$\nu=\frac{1}{2m+1}$ with $m$ an integer. It is noted that
according to this formalism for a quantum Hall particle the charge
is taken to be given by $-\nu e$ when $\nu$ is the filling factor.

As $\mu$ here corresponds to the monopole strength, we can relate
this with the Berry phase. Indeed $\mu=\frac{1}{2}$ corresponds to
one flux quantum and the flux through the sphere when there is a
monopole of strength $\mu$ at the centre is $2\mu$ . The Berry
phase of a fermion of charge $q$ is given by $e^{i\phi_B}$ with
$\phi_B=2\pi q N$ where $N$ is the number of flux quanta enclosed
by the loop traversed by the particle.

If $\mu$ is an integer, we can have a relation of the form
\begin{equation}
{\bf J} ~=~ {\bf r} \times {\bf p} - \mu {\bf \hat{r}} ~=~ {\bf
r}^{~\prime} \times {\bf p}^{~\prime}
\end{equation}
which indicates that the Berry phase associated with $\mu$ may be
unitarily removed to the dynamical phase. Evidently, the average
magnetic field may be considered to be vanishing in these states.
 The attachment of
$2m$ vortices ($m$ an integer) to an electron effectively leads to
the removal of Berry phase to the dynamical phase. So, FQH states
with  $2 \mu_{eff} = 2m + 1$  can be viewed as if one vortex is
attached to the  electron. Now we note that for a higher Landau
level we can consider the Dirac quantization condition $e
\mu_{eff} = \frac{1}{2} n$, with $n$ being a vortex of strength $2
\ell + 1$. This can generate FQH states having the filling factor
of the form $\frac{n}{2 \mu_{eff}}$ where both $n$ and $2
\mu_{eff}$ are odd integers. Indeed, we can write the filling
factor as \cite{pb1,pb2}
\begin{equation}
\nu ~=~\frac{n}{2 \mu_{eff}} ~=~ \frac{n}{(2 \mu_{eff}~ \mp 1) \pm
1}  ~=~\frac{n}{2m' \pm 1}~=~ \frac{n}{2mn \pm 1} \label{nu1}
\end{equation}
where $2 \mu_{eff} \mp 1$ is an even integer given by $2m'=2mn$.
The particle-hole conjugate state can be generated with the
filling factor given by
\begin{equation} \nu ~=~ 1 - \frac{n}{2mn \pm 1}
~=~ \frac{n (2m - 1) \pm 1}{2mn \pm
1}=\frac{n^{\prime}}{2mn^{\prime} \pm 1} \label{e3}
\end{equation}
where $n(n')$ is an odd(even) integer.

 Recently, some novel
generation of filling factors for FQH fluid have been observed
\cite{pan} which do not satisfy the primary Jain sequence. An
analysis of these states from a Berry phase approach \cite{novel}
suggests that the observed filling factors like $\nu=4/11$,
$5/13$, $6/17$, $4/13$ and $5/17$ corresponds to the lowest Landau
level and are given by
$$\nu=\frac{n}{2\mu_{eff}}=\frac{n}{2m'\pm 1}$$
where $m'$ is such that it cannot be split into the form $mn$. The
FQH state $\nu=7/11$ is found to be particle-hole conjugate state.

Besides, there are some FQH states with even denominator filling
factors which also needs explanation. Indeed in some earlier
papers \cite{bpeven, bdp}, it has been pointed out that FQH states
at even denominator filling factor appear as pairs and correspond
to non-Abelian Berry phase. From our previous discussions we note
that the Dirac quantization condition $e\mu=1/2$ suggests that 
$\nu=1/2$ FQH state corresponds to the factor $\mu=1$. However,
from the angular momentum relation (4) we note that in case $\mu$
is an integer the Berry phase may be unitarily removed to the
dynamical phase. This implies that the average magnetic field may
be taken to be vanishing in these states. However, the effect of
the Berry phase may be observed when we split the state into a
pair with each component having the constraint of representing the
state with $\mu=1/2$. This corresponds to non-Abelian Berry phase
for these states.

\section{Statistics of Quasiparticles in FQH states}
In an earlier paper \cite{hajra}, it has been pointed out that the
quantization of a Fermi field is achieved when we introduce an
anisotropy in the internal space through the introduction of a
$direction$ $vector$ attached to the space-time point. Indeed, to
have quantization in Minkowski space we have to take into account
a complex manifold when the coordinate is given by $z^\mu =x^\mu
+i \xi^\mu$ where $ \xi^\mu$  is the $direction$ $vector$ attached
to the space-time point $x^\mu$ \cite{kp}. The two opposite
orientations of the $direction$ $vector$ correspond to particle
and anti-particle states. This $direction$ $vector$ corresponds to
the internal helicity when $\xi^\mu$ is written in terms of the
spinorial variable $\xi^\mu ={\lambda^\mu}_{\alpha} \theta^\alpha$
($\alpha=1,2 $) where $\theta$ is a two component spinor. The
corresponding metric $g_{\mu\nu} (x,\theta,{\bar{\theta}})$
eventually gives rise to a gauge theoretic extension when the
position and momentum variables can be written as \cite{hajra}
\begin{equation}\label{var}
  Q_\mu= -i (\frac{\partial}{\partial p_\mu} + B_\mu)~~~~~
 P_\mu= i (\frac{\partial}{\partial q_\mu} + C_\mu)
\end{equation}
where $B_\mu(C_\mu)$ corresponds to a $SL(2,C)$ gauge field. Here
$q_\mu(p_\mu)$ denotes the mean position (momentum) in the
external observable space. In fact, the $direction$ $vector$
effectively represents a vortex line which is topologically
equivalent to a magnetic flux quantum and the gauge field
theoretical extension corresponds to the magnetic field associated
with the internal extension of the particle \cite{royp}. This
leads to the description of a massive fermion as a Skyrmion.

When the $direction$ $vector$ $\xi^\mu$ is attached to the
space-time point $x^\mu$, the corresponding field function is
given by $\phi(x^\mu,\xi^\mu)$ which can be treated to describe a
particle moving in an anisotropic space and the wave function
takes into account the polar coordinate $r,\theta, \phi$ for the
space component of the vector $x^\mu$ along with the angle $\chi$
specifying the rotational orientation around the $direction$
$vector$. The eigenvalue of the operator
$i\frac{\partial}{\partial \chi}$ just corresponds to the internal
helicity associated with the vortex line (magnetic flux quantum).
Indeed, the angular momentum operator is now given by
\begin{equation}\label{ang}
  {\bf J}={\bf r} \times {\bf p}- \mu \hat{{\bf r}}~~~\mu=0,\pm
  1/2, \pm 1, \pm 3/2....
\end{equation}
which is similar to that of a charged particle moving in the field
of a magnetic monopole of strength $\mu$.

The spherical harmonics incorporating the term $\mu$ is given by
\cite{fierz, hurst}
\begin{equation}\label{sphe}
  {Y_l}^{\mu,\nu}=(1+x)^{-(m-\mu)/2} (1-x)^{-(m+\mu)/2}
  \frac{d^{l+m}}{d^{l-m}x}\left[
  (1+x)^{l-\mu}(1-x)^{l+\mu}\right]e^{im\phi}e^{-i\mu\chi}
\end{equation}
where $x=cos\theta$. Here $m$ and $\mu$ represents the eigenvalues
of the operators $i\frac{\partial}{\partial \phi}$ and
$i\frac{\partial}{\partial \phi}$ respectively.

 In this formalism, a fermion is represented by a boson attached
 with a magnetic flux quantum. The Berry phase acquired by the
 boson when traversing a loop enclosing the flux quantum is given
 by $e^{i2\pi\mu}$ \cite{dp}. Indeed, the angular part associated
 with the angle $\chi$ in the spherical harmonics ${Y_l}^{m,\mu}$
 is given by $e^{-i\mu\chi}$ where we have
\begin{equation}\label{ope}
  i\frac{\partial}{\partial\chi}=\mu e^{-i\mu\chi}
\end{equation}
Thus when $\chi$ is changed to $\chi+ \delta\chi$ we have
\begin{equation}
i \frac{\partial}{\partial (\chi + \delta \chi)} ~ e^{- i \mu
\chi} ~=~ i \frac{\partial}{\partial (\chi + \delta \chi)} ~ e^{-
i \mu (\chi + \delta \chi)} ~ e^{i \mu \delta \chi}
\end{equation}
which implies that the wave function will acquire an extra factor
$e^{i \mu \delta \chi}$ due to infinitesimal change of the angle
$\chi$. As the angle $\chi$ is changed over the closed path $0
\leq \chi \leq 2 \pi$, for one such complete rotation, the wave
function will acquire, the required phase
\begin{equation}
 \exp~\left\{  i \mu ~ \int^{2 \pi}_{0} ~ \delta \chi \right\} ~=~
e^{i 2 \pi \mu}
\end{equation}

As $\mu=1/2$ corresponds to one flux quantum, when a boson
traverses a closed path enclosing one flux quantum we have the
phase $e^{i \pi}$ which suggests that the system represents a
fermion implying that the wave function changes sign after $2\pi$
rotation. The effect of the flux tube is to induce an appropriate
Aharanov-Bohm phase which simulates the statistical phase factor.
Indeed, as a fermion is described by a scalar particle carrying
one flux quantum, we note that when a fermion traverses a loop
enclosing another fermion we can view it as if a boson traverses a
loop enclosing two magnetic flux quanta so that the Berry phase is
given by $e^{i2\pi \mu}$ with $\mu=1$ $i.e.$ $e^{i2\pi}$. This
implies that when a fermion encloses $N$ number of flux quanta the
Berry phase is $2\pi N$ \cite{pbklu}. Now a process which
exchanges two fermions can be viewed as if one of the fermions
moving about the other in a half circle. Thus the statistical
phase $e^{i\pi\theta}$ is given by $\theta=1$. This is the
statistical phase factor when two fermions are adiabatically
exchanged. This suggests that when a boson (scalar particle) is
dressed with $n$ vortices (magnetic flux quanta) the statistical
phase is given by $e^{i\pi\theta}$ with $\theta=n$.

Now to study the statistics parameter for FQH states, we note that
according to the formalism as described in the previous section,
the FQH states at $\nu=\frac{1}{2m+1}$ corresponds to the fact
that an electron is attached with $(2m+1)$ vortices of which $2m$
vortices contribute to the dynamical phase. Now taking into
account the duality principle \cite{fish} which interchanges the
role of charges and vortices we may consider vortices as bosons
which see each original particle as a flux tube carrying one Dirac
flux quantum. That is, an electron attached with $m$ vortices may
be viewed as if $m$ hard-core bosons share one vortex (magnetic
flux quantum). So quasiparticles describing FQH states at
$\nu=\frac{1}{2m+1}$ may be viewed as such that $(2m+1)$ hard-core
boson are attached with one flux quantum implying that each
hard-core boson shares the vortex density $\frac{1}{2m+1}$. So
from our above analysis we note that the statistical phase will be
given by $e^{i\pi\theta}$ with $\theta=\frac{1}{2m+1}$. So for the
FQH state at $\nu=1/3$ and $1/5$, we will have the statistics
parameter $\theta=1/3$ and $1/5$ respectively. As under the
duality principle, the charge is now shared by the hard-core
bosons the charge of the quasiparticle will be given by $-\nu
e$,$i.e.$ for $\nu=1/3$ and $1/5$, the charge will be $-e/3$ and
and $-e/5$ respectively. This also follows from our analysis in
sec. 2 where we have pointed out that from the Dirac quantization
condition $e\mu=1/2$, with $\mu=3/2(5/2)$ we obtain $\nu=1/3(1/5)$
implying that the charge is $-\nu e$.

To consider the statistics parameter for FQH states at $\nu
=\frac{n}{2mn \pm 1}$ with $n>1$ and odd, we have noted in the
earlier section that this corresponds to the higher Landau level
which follows from the Dirac quantization condition
$e\mu=\frac{1}{2}n$, where $n$ corresponds to the vortex of
strength $2l+1$. This implies that under the duality principle,
$n$ units of flux density $\frac{1}{2mn \pm 1}$ are attached with
each hard core boson. So from our above analysis, we find that the
statistical phase $e^{i\pi\theta}$ will be given by
$\theta=\frac{n}{2mn \pm 1}$. Also the charge will be given by
$-\nu e$. This suggests that for the FQH states at $\nu=3/5, 3/7,
5/9 ...$ we have $\theta~=3/5,~ 3/7,~ 5/9 $ and so on.

For FQH states at $\nu=\frac{n}{2mn' \pm 1}$ with n an even
integer, we discussed in sec. 2 that these corresponds to
particle-hole conjugate states. Indeed, these are given by the
relation $$\nu~=~1-\frac{n}{2mn \pm 1}=\frac{n'}{2mn' \pm 1}$$
where $n(n')$ is an odd (even) integer. So, under the duality
principle, the effective flux density associated with each
hard-core boson will be $1-\frac{n}{2mn \pm 1}$, such that the
statistical phase $e^{i\pi\theta}$ is given by $\theta
=1-\frac{n}{2mn \pm 1}=\frac{n'}{2mn' \pm 1}$. Thus we find that
the statistical phase $e^{i\pi\theta}$ for a quasiparticle in FQH
states at filling factor $\nu=\frac{n}{2mn \pm 1}$, $n$ odd or
even corresponds to $\theta=\nu=\frac{n}{2mn \pm 1}$ and charge is
$-\nu e$. For the quasihole, the statistics parameter will be
identical to that of a quasiparticle and the charge will be of
opposite sign.

The above result for statistics parameter can be generalized to
the newly discovered states also \cite{pan} and we will have the
statistical phase given by $e^{i\pi\theta}$ with $\theta=\nu$ and
the charge for the quasiparticles (quasiholes) is $-\nu e (\nu
e)$.

This analysis can be generalized to study the statistics parameter
for the FQH states with even denominator filling factor. In this
case, we observe that for FQH state at $\nu=1/2$, as this
corresponds to $\mu=1$ the quasiparticle corresponding to the pair
state effectively represent that this is dressed with two
vortices. Indeed, the quasiparticle for FQH liquid at $\nu=1/2$
corresponds to a singlet characterized by a flux
$\phi_0=\frac{hc}{2e}$. So under the duality principle, this
implies that two hard-core bosons share one Dirac flux quantum. So
from our above analysis we find that the statistical phase
$e^{i\pi\theta}$ is given by $\theta=1/2$. This also implies that
the charge of the quasiparticle will be given by $-\frac{1}{2}e$.
For the Haldane-Rezayi FQH state at $\nu=5/2$, it has been
observed that this corresponds to higher Landau level ($l=2$). The
Dirac quantization condition $e\mu=n/2$ with $n=2l+1$ suggests
that for $l=2$ we have $\nu=5/2$ FQH state. Generalizing our above
analysis we note that under duality principle this implies that
$5$ units of flux density $1/2$ are attached with each hard-core
boson and as such the statistical phase $e^{i\pi\theta}$will be
here characterized by the parameter $\theta=5/2$. Following our
above argument we note that the charge here will also be given by
$-\frac{5}{2}e$. The same argument suggests that the newly
observed even denominator FQH state such as $\nu=3/8 $ and
$\nu=3/10$ \cite{pan} will also have the statistical phase
$e^{i\pi\theta}$ with $\theta=3/8$ and $\theta=3/10$ respectively.
\section{Discussion}
Our above analysis suggests that we have a generalized relation
such that the statistical phase of a quasiparticle (quasihole) for
FQH states at $\nu=p/q$ with $p(q)$ even or odd is given by
$e^{i\pi\theta}$ with $\theta=p/q$ and the charge for a
quasiparticle is $-\frac{p}{q}e$ whereas for a quasihole the
charge will be of opposite sign. This is consistent with that
observed in numerical studies with the Laughlin's wave function
for the quasiholes at $\nu=1/3$ though for quasiparticles the
result is not straight forward for finite number of electrons and
such that it may possibly converge to $1/3$. This discrepancy is
due to the fact that for Laughlin wave functions the inverse of
the operator used to create a quasielectron is not simply the
conjugate of the inverse quasihole operator. In the composite
fermion (CF) scenario a general analysis suggests that for the FQH
state at $\nu=~\frac{n}{2pn+1}$ the statistics parameter is
$\theta=~\frac{2p}{2pn+1}$\cite{jeon}. However numerical studies
\cite{lei,jeon} have suggested that $\theta$ has negative sign.
Jeon et al \cite{jeon} have conjectured that the sign enigma is
due to a small perturbation caused by the insertion of an
additional CF quasiparticle. Evidently our above result for
$\theta$ is at variance with the prediction of the CF formalism.

Based on the hierarchical model, Halperin \cite{halperin} has
derived a recurrence relation for the statistical angle of the
elementary charged excitations. This recurrence relation is found
to be equivalent to the explicit analysis of Su \cite{23} based on
adiabatic theorem. Though the results obtained in Ref.[23] are
consistent with our results for values at $\nu=1/m$, it is not so
for other states. The discrepancy has its origin in the fact that
unlike the hierarchical model, in our formalism quasiparticles are
not considered as composites of elementary excitations.

In the present framework we note that quasiparticles (quasiholes)
of FQH states are anyons when projected in $2D$ but these anyons
are not ideal point particles. Indeed these are extended objects
when a magnetic flux density is dressed with a hard core boson. It
may be mentioned here that in some recent papers\cite{noncom,pg}we
have pointed out that the quasiparticles at $\nu=1/m$ may be
considered as underlying fields in noncommutative manifold $M_4
\times Z_N$ with $N>2$ and odd. In fact it is pointed out that
while the noncommutative manifold  $M_4 \times Z_2$  has its
underlying field as fermion, the manifold  $M_4 \times Z_N$  with
$N>2$ and odd has its underlying fields with fractional statistics
having the statistics parameter $\theta=1/N$. The discrete space
when considered as the internal space the system corresponds to an
extended body dressed with magnetic flux density.

\end{document}